# The manifestations of 'l-Doubling' in gas-phase rotational dynamics

Kfir Rutman Moshe*[1,2], Dina Rosenberg*[1,2], Inbar Sternbach, and Sharly Fleischer[1,2†]

[1]Raymond and Beverly Sackler Faculty of Exact Sciences, School of Chemistry, Tel Aviv University 6997801, Israel.
[2]Tel-Aviv University center for Light-Matter-Interaction, Tel Aviv 6997801, Israel
[†]Email: sharlyf@tauex.tau.ac.il
* Equal contribution

**Abstract**

The 'l-Doubling' phenomenon emanates from the coupling between molecular rotations and perpendicular vibrations (bending modes) in polyatomic molecules. This elusive phenomenon has been largely discarded in laser-induced molecular alignment. Here we explore and unveil the ramifications of 'l-Doubling' to the coherent rotational dynamics of triatomic molecules at ambient temperatures and above. The observed 'l-Doubling' dynamics may be wrongly considered as collisional decay throughout the first few hundreds of picoseconds past excitation, highlighting the importance of correct assimilation of l-Doubling in current research of dissipative rotational dynamics and in coherent rotational dynamics in general.

**Introduction**

Upon their interaction with gas-phase molecules, ultrashort laser pulses impart short duration torque which manifest as coherent molecular rotations. This scenario has provided a rich playground for various scientific endeavors over the last 3-4 decades [1–4]. The initial motivation of the field aimed at obtaining accurate rotational coefficients and associated molecular structure[5–9] titled 'rotational coherence spectroscopy'. The latter has evolved into coherent rotational control ('molecular alignment') aiming to transiently lift the inherent isotropic angular distribution of gas ensembles and facilitate 'molecular frame spectroscopies' via advanced optical techniques such as HHG [10], Coulomb explosion imaging [11], ultrashort x-ray diffraction[12] and various others. In recent years the interest has gradually shifted toward the underlying processes that govern decay and decoherence phenomena in long-lasting coherent rotational dynamics. Within this realm, advanced experimental methods have been developed in the context of rotational echo spectroscopies[13,14] and advanced modeling [15–17], that shed new light on the quantum mechanical aspects of rotating molecular ensembles [18–22].

In this work we present and explore the ramifications of 'l-Doubling' on the dynamics of coherently rotating linear polyatomic molecules as revealed in molecular alignment experiments. The phenomenon was first described by Hertzberg in 1942 [23] to explain the energy splitting observed in ro-vibrational spectroscopy of linear polyatomic molecules[24]. In his paper, Hertzberg coined the term 'l-Doubling' in analogy to the Λ-doubling phenomena in electronic spectra[25], and attributed the effect primarily to the displaced position of the perpendicular vibration, which renders linear polyatomic molecules slightly asymmetric. While Herzberg mentioned that 'l-Doubling' results, to some extent, from Coriolis coupling of different vibrations, his paper initiated a long-lasting altercation with Nielsen and Shaffer who attributed l-Doubling solely to the Coriolis coupling[26]. Without delving into the controversy that was beautifully detailed by James K.G. Watson [27], the accepted theory was given by Nielsen in 1951 [28] who derived the expression for l-Doubling from Coriolis coupling solely.

In what follows we provide a simplified explanation for the l-Doubling phenomenon that manifests as rotational energy splitting due to coupling with the bending modes of a linear triatomic molecules. For brevity, let us consider a molecule restricted to rotate in the xy plane. At the ground vibrational level, the rotational constant of this molecule is given by $B_{000}$. Perpendicular vibration (bending) of the molecule can occur in all directions occupying the yz plane, spanned by the two axial bending modes: bending along the y-axis and bending along the z-axis. Figures 1a,b depict 5 representative molecular geometries assumed by the molecule respectively.

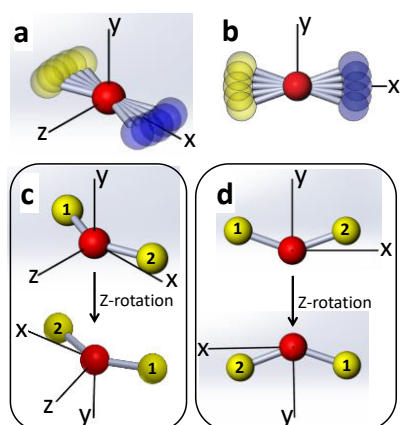

Figure 1: Graphical representation of molecular geometries of the two axial bending modes: (a) along the z-axis and (b) along the y-axis. Bend angles are exaggerated for better visibility.
(c,d) symmetry association of the two bending modes upon rotation about the z-axis. Relevant for homonuclear triatomic molecules – see symmetry considerations in the text of the $CO_2$.

Since the bending frequency ($\sim 500 cm^{-1}$) is much larger than the rotational frequency ($\leq 1 cm^{-1}$), we consider an effective moment of inertia for each axis by averaging over all of the geometries assumed by the molecule throughout one bending period. This results in three distinct inertial moments:

$I_x$ (which governs the rotation about the long molecular axis) is slightly shifted from its ground state value (zero) yet remains negligibly small.

$I_y$ and $I_z$ are both shifted to slightly lower values and their rotational constants slightly increased respectively. Most importantly, due to the bending of the molecule, the degeneracy of $I_y$ and $I_z$ is lifted and two rotational constants are formed, distinct both by their value and rotational symmetry (parity):

$B(\Pi^-) = B_0 - \alpha_t + \frac{1}{2}\bar{q}_t$ and $B(\Pi^+) = B_0 - \alpha_t - \frac{1}{2}\bar{q}_t$.

With $\bar{q}_t = \frac{2B_e^2}{\omega_t}\left[1 + 4\omega_t^2 \sum_s \frac{\xi_{st}^2}{\omega_s^2 - \omega_t^2}\right]$ where $B_e$ is the equilibrium rotational constant, $\xi_{st}$ is a dimensionless Coriolis coupling coefficient that is determined by the coordinates of the stretching (indexed by $s$) and bending (indexed by $t$) vibrational modes. We refer the bold readers to the complete derivation by Nielsen in ref.[28] where the constant $\alpha_t$ may be traced. In what follows we humbly proceed to the experimental manifestations of l-Doubling and corresponding simulations utilizing the experimental values of the rotational constants obtained previously (see table of in supplementary information section 1).

**Experimental**

We perform time-resolved optical birefringence measurement of the anisotropic angular distributions of rotationally excited gas throughout its long field-free dynamics. Our setup uses the 'weak field polarization detection' method [29,30] in collinear configuration [14,18,31–33] to increase the interaction length between the pump and probe within the gas cell to enable measurements of gas samples in the single torr level. A femtosecond (fs) laser beam (~110fs, 800nm) from a Ti:Sapphire Chirped Pulse Amplifier is split to form pump (90%) and probe (10%) that is frequency doubled in a BBO crystal to form a weak 400nm probe. The timing of the probe is computer controlled using a long (50cm, ~1.6ns) delay stage. The polarization of the 400nm probe is set at 45° to the 800nm pump beam, and the two are combined by a dichroic mirror as in our previous work [18]. Here, however the pump (800nm) and probe (400nm) beams are focused by two selective lenses mounted on a linear stage to enable further optimization of the overlap between the beams within the gas cell. The latter has further improved the S/N of the measurement by ~3 fold and provided clear signals from gas densities of the level of single torr. The 800nm pump beam is filtered out by a short-pass filter (BG40) and the transmitted 400nm probe is analyzed for changes in its polarization by a λ/4 plate, a Wollaston prism and a pair of balanced photodetectors.

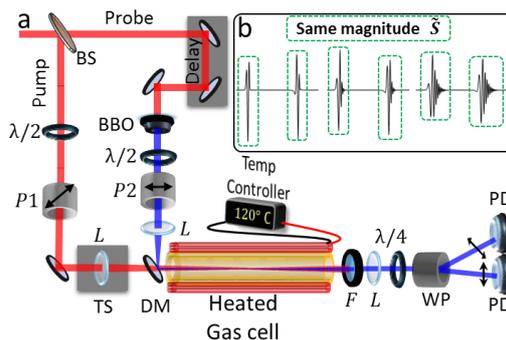

Figure 2: Transient birefringence setup. $P$ – polarizer, L- lens, λ/2-half wave plate, λ/4 – quarter wave plate, BBO crystal, WP- Wollaston prism. In the upper right corner we plot several alignment transients exemplifying the effect of centrifugal distortion. Using the quantification method ($\tilde{S}$, see text) we obtain the same magnitude for all these transients.

## l-Doubling manifestation in rotational decay dynamics

Figure 3 exemplifies the ramifications of l-Doubling by comparing between the rotational decay dynamics of methyl-iodide ($CH_3I$, Fig.3a) and carbonyl sulfide (OCS, Fig.3b).

Each data point in Fig.3 quantifies a specific 'revival signal magnitude' extracted from the time-resolved birefringence of the gas. We use a quantification metric that is immune to the centrifugal distortion dephasing used in our previous work [18,34] and exemplified in Fig.2b where all of the alignment transients (each surrounded by dashed green line) have the exact same magnitude $\tilde{S}$, regardless of their different shapes. This is done by integrating over each revival transient squared $\tilde{S} = \int_{t_{rev}-\varepsilon}^{t_{rev}+\varepsilon}|S_{rev}(t)|^2 dt$ or alternatively by integrating over the power spectrum of each revival: $\tilde{S} = \int_{\omega_1}^{\omega_2}|\mathcal{F}[S_{rev}(t)]|d\omega$. The latter is used in this work where the integration is performed over the range $[5cm^{-1}, 80cm^{-1}]$.

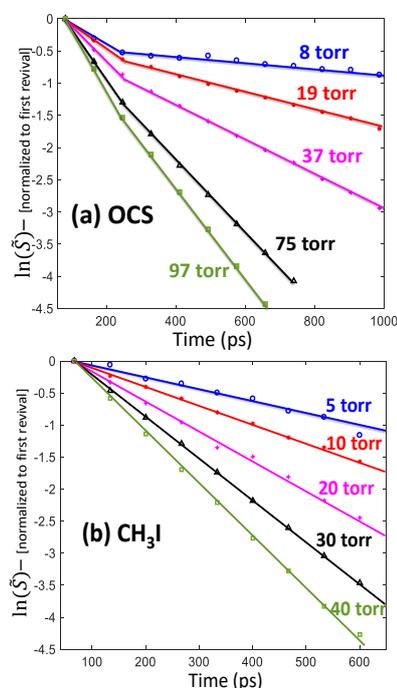

Figure 3: Experimental alignment magnitude ($\tilde{S}$) for (a) $CH_3I$ and (b) OCS with different gas pressures at ambient temperature.

Each gas pressure is color coded in Fig.3 and presented on a semi-logarithmic scale. In $CH_3I$, the slope is purely linear (single exponential) and increases with gas pressure as expected from collision-induced decay. In OCS however, we find the decay to follow a double exponential decay as evident by the two distinct decay slopes where the decay rate through the first three revival periods (up to ~$250ps$ past excitation) significantly exceeds the collisional decay observed at longer times. While the different slopes of the two fitted lines are unmissable at gas pressures of 8, 19 and 37torr, they tend to coincide as the pressure increases and at 97 torr (green data set in Fig.3b) they are hardly decipherable. At higher gas pressures (not shown here) the dynamics can be erroneously fitted excellently by a single exponential decay as in CH$_3$I. The elusiveness of the l-Doubling phenomenon under investigation here can be appreciated by the results of Fig.3 as it is practically overshadowed by collisional decay already at sub-atmospheric pressures.

**l-Doubling in OCS at Different Temperatures**

In order to decipher the dynamics imposed by l-Doubling from the overshadowing collisional decay, we set the experimental conditions to optimize the visibility of the former and minimize the latter. This is done by increasing the thermal population of the excited bending mode and by decreasing the gas pressure (5 torr OCS in Fig.4) respectively. Furthermore, owing to their selective temporal signatures as shown in Fig.3a, one can decipher between the l-Doubling and collisional decay phenomena. While the first few revival periods exhibit the effects of both the l-Doubling and collisional decay, at longer delays the dynamics is governed solely by collisions. Thus, by quantifying the collisional decay rate in the region of $t > 800ps$, we can filter out the collisional decay and effectively isolate the l-Doubling effect for the three gas temperatures depicted in Fig.4 (293K - blue, 333K – green, and red - 393K). First, we calculated the magnitude of each revival transient ($\tilde{S}$) as noted above. Then we fitted the calculated magnitudes in the region $t > 800ps$ to a single exponent $\tilde{S}_{10T_{rev}} \cdot e^{-\gamma t}$ (where $t$ is in units of $T_{rev} = (2Bc)^{-1}$) and extracted the collisional decay rates ($\gamma = 6.5 \cdot 10^{-4}, 6.7 \cdot 10^{-4}$ and $7.2 \cdot 10^{-4}$ in units of $T_{rev}^{-1}$) for the three gas temperatures $293K, 333K$ and $393K$ respectively).

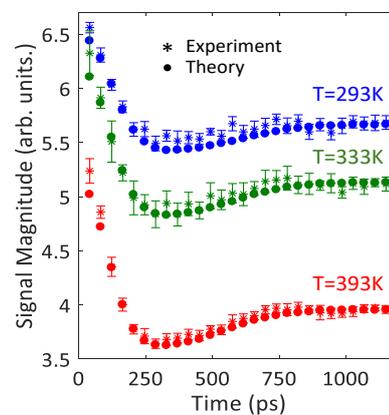

Figure 4: Revival signal magnitudes ($\tilde{S}$) of $OCS$ (5 torr gas pressure) at 293K (blue), 333K (green) and 393K (red). The exponential collision-decay was extracted from the plateau region and filtered out (see text).

The decay-free revival magnitudes ($\tilde{S}_{exp}$) were obtained by dividing the revival magnitudes by $e^{-\gamma t}$ of OCS and are marked by the color-coded asterisk symbol overlaid with the numerically simulated color-coded points.

The simulations were performed using the density matrix formalism as in our previous works [18,35]. We numerically propagate the Liouville-Von Neumann equation, $\frac{d\rho}{dt} = -\frac{i}{\hbar}[\hat{H}, \rho]$, where $\hat{H} = \frac{\hat{L}}{2I} + V(\theta, t)$ is the Hamiltonian, $\hat{L}$ is the angular momentum operator, $I$ is the moment of inertia and $V(\theta, t) = -\frac{1}{4}\Delta\alpha|E(t)|^2 cos^2\theta$ is the interaction term for nonresonant rotational excitation with $\Delta\alpha$ the anisotropic polarizability of the molecule. Here however, we include the rotational contributions of molecules populating the excited vibrational states in addition to the ground state. Since the interaction term does not couple between the rotational state manifolds of different vibrational states, we can calculate the pulse-induced rotational dynamics associated with each populated vibrational level separately, and sum the results weighted by their Boltzmann factor. For OCS, with $\omega_{bend} = 539 cm^{-1}$ we used the rotational coefficients $B_{000} = 0.20285 cm^{-1}$, $B_{010}^+ = 0.20331 cm^{-1}$ and $B_{010}^- = 0.20310 cm^{-1}$ for the ground, excited in-plane and excited out-of-plane vibrations respectively [36] with their corresponding Boltzmann factors at the different temperatures. The simulated alignment signal was convolved with a 350fs Gaussian filter to account for the length of the probe pulse measured separately in CCl$_4$ reference gas (isotropic polarizability) and the signal magnitudes calculated ($\tilde{S}_{sim}$). The simulated magnitudes were factored by the ratio of the experimental and simulated plateau region ($t > 800ps$) and resulted in very good agreement with the experimental data.

Fig.4 shows that l-Doubling manifests initially as fast decay in signal magnitude at $0 \rightarrow 300ps$ followed by partial recovery at $300ps \rightarrow 800ps$. This is merely the beating of different revival signals with slightly different rotational coefficients associated with three vibrational levels. Shortly after the excitation, at $1/2\, T_{rev} = (4Bc)^{-1} \sim 41ps$, all three alignment signals are temporally synchronized hence they interfere constructively to yield the maximal signal magnitude. As time proceeds, they gradually de-synchronize and at $\sim 300ps$ ($\sim 3.5 T_{rev}$) they interfere destructively and yield the minimal alignment magnitude. As time propagates the temporal overlap of the three contributing signals further reduces until complete temporal separation is achieved at $\sim 800ps$ and the magnitude of the (collision-free) signal remains constant. The modulation depth, namely the ratio between the signal dip at $\sim 300ps$ and the first signal at $\sim 41ps$ depends on the relative amplitudes of the interfering signal contributions, and increase with the gas temperature. At room temperature (293K) this amounts to 16% decrease in signal magnitude, while at 333K and 393K the signal magnitude decreases by 21% and 27% respectively owing to the increased thermal population of the excited bending level.

### **l-Doubling in symmetric molecules: the case of CO$_2$**

Symmetric molecules (e.g. N$_2$, CO$_2$, C$_2$H$_2$ etc.) obey the Pauli principle which states that the total wavefunction of the molecule must be either symmetric or anti-symmetric upon exchange of their identical Bosonic or Fermionic nuclei respectively. At the ground electronic and vibrational states this implies direct symmetry relations between the nuclear spin and rotational wavefunctions' symmetries [37] and enable selective rotational manipulations of Ortho and Para molecular spin isomers[38,39]. Correspondingly, CO$_2$ molecules at the ground electronic and vibrational states populate only symmetric rotational states with even J quantum numbers (J=0,2,4,…2n) owing to the Bosonic nuclear spin of the Oxygen atoms ($I_{O^{16}} = 0$) [40]. The symmetry of the vibrational levels imposes additional symmetry considerations, the manifestation of which in the l-Doubling dynamics is discussed and demonstrated in what follows.

In Figs.1a,b we used a non-symmetric triatomic model to visualize the distinct bending modes and their lifted degeneracy upon rotation, resulting in distinct $B(\Pi^-)$ and $B(\Pi^+)$ rotational coefficients. In Figs.1c,d we switch to a symmetric triatomic (with identical nuclei, marked by '1' and '2'), where in addition to the lifted rotational degeneracy, the two bending modes are associated with opposite rotational symmetries. To visualize these symmetry considerations, we freeze the molecular geometry at the turning point where the molecule is maximally bent ('V' shaped) and exchange between the two nuclei by rotation about the z-axis. The opposite symmetries are readily observed: while the z-bend (Fig.1c) is symmetric upon z-rotation, the y-bend (Fig.1d) is anti-symmetric upon z-rotation. Correspondingly, the z-bend is associated with purely symmetric rotational levels (Even J's) and the y-bend populates only anti-symmetric rotational levels (Odd J's). Due to their different rotational dynamics at $\frac{1}{4}, \frac{3}{4} T_{rev}$ (and recurrences) [1,41,42] the l-Doubling manifests as partial alternations between adjacent signal magnitudes in addition to the temporally long beating dynamics.

Figure 5 shows the experimental magnitudes of $CO_2$ gas sample at 393K overlaid with the simulated results (the collisional decay was filtered out as discussed before). The half-integer revival signals ($\frac{n}{2} T_{rev}$) are depicted in blue and the quarter-revival signals ($\frac{n}{2} + \frac{1}{4} T_{rev}$) in red, demonstrating distinctively different dynamics. The data is normalized to the magnitude of the first blue signal ($t = \frac{1}{2} T_{rev}$) for brevity. The trend of the blue signals resembles that of OCS (Fig.4) starting with rapid decrease in magnitude, followed by partial recovery to the level of $\sim 0.85$ at the plateau region ($t > \sim 800 ps$). The red signal however, starts at the plateau level of 0.85 and rapidly decreases to $\sim 0.78$ (at $t = 180 ps$) after which the trend reverses and the signal increases to $\sim 0.89$ at $t = 440 ps$ (above its initial level) and then decreases back to the plateau. These dynamics result from interference of the rotational contributions associated with the ground vibrational state ($B_{000} = 0.3902 cm^{-1}$) and with the excited vibrational states ($B_{010}^+ = 0.3912 cm^{-1}$ and $B_{010}^- = 0.3905 cm^{-1}$) [43]. For $\omega_{bend} = 667 cm^{-1}$ and temperature of $393 K$ the Boltzmann populations are 85.2%, 7.4%, 7.4% respectively.

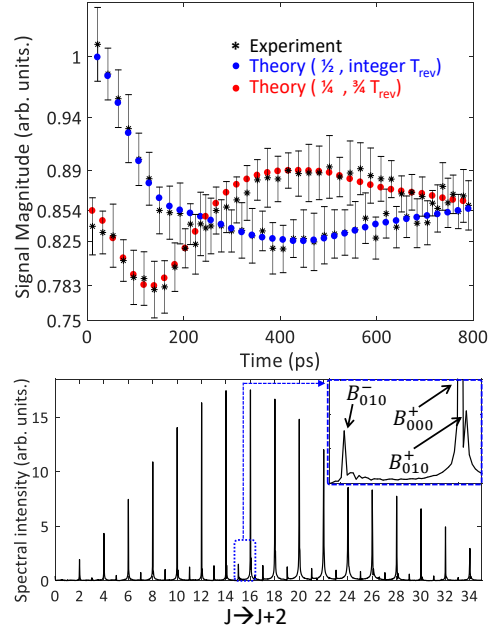

Figure 5: (a) Alignment magnitudes of $CO_2$ gas (7torr, 393K). The experimental data is given by the black asterisks overlaid with the simulated magnitudes (blue dots – half-integer revivals, red dots – quarter revivals). Note that the simulated time-dependent alignment was convolved with a 350fs Gaussian to account for the experimental probe length (measured separately in $CCl_4$ gas sample). (b) Spectral intensity of the J→J+2 transition lines extracted from Fourier transformation of the raw time time-resolved alignment signal. The inset shows the 15→17 and 16→18 transitions where the three contributing vibrational levels are marked.

In order to understand the dynamics of Fig.5 we note that the transient alignment signals coming from odd and even rotational states are $\pi$-phase shifted at quarter revivals (red signals) and identical otherwise (red signals) [38]. In vicinity to the time of excitation, all three contributions are temporally synchronized, and interfere constructively at $t = \frac{1}{2} T_{rev}$ (the first blue signal with magnitude normalized to 1). At $t = \frac{1}{4} T_{rev}$ (first red signal), the contributions of $B_{010}^+ (even\ J's)$ and $B_{010}^- (odd\ J's)$ are also synchronized, but owing to their inherent $\pi$-phase they interfere destructively, leaving the $B_{000}(even\ J's)$ contribution solely (with magnitude 0.85). As time progresses, the three signal components gradually de-synchronize owing to their different rotational constants and the overall signal magnitude decreases. The exact time at which the blue and red signals reach their minimal and maximal magnitudes (here around $t = 440 ps$) depends on the difference in rotational constants as well on the transient signal duration. The latter is demonstrated in section 2 of the supplementary information. We note that the minimal magnitude of the blue (0.825) and maximal magnitude of the red (0.89) are equally distant from the plateau magnitude (0.854) due to partial temporal overlap among the three signal components that manifest in destructive and constructive interference respectively. When the three signal components become separated in time, they stop interfering with each other and the magnitude settles to the constant plateau value of 0.854.

To complete the discussion we present the frequency domain in Fig.5b, obtained by Fourier transformation of the time-resolved alignment signal (not shown here). We plot the spectral intensity as a function of the Raman transition between J and J+2 quantum states of the ground vibrational level using: $\widetilde{\omega}_{J \to J+2} = E_{J+2} - E_J = B_{000}^{+}(4J + 6)$ where we discarded the centrifugal distortion for brevity. The highest populated states (high intensity peaks) are associated with the ground vibrational state (with $B_{000}^{+}$) and occupy the even J→J+2 transitions. The odd J→J+2 transitions are attributed to the excited vibrational level ($B_{010}^{-}$) with low intensities (~7.4% population noted above). The inset shows an enlarged view of the 15→17 and 16→18 transitions where the latter shows the $B_{010}^{+}$ transition to the right of the ground state peak (marked in the inset).

In order to further verify our theoretical and experimental strategy for selectively extracting the l-Doubling dynamics presented throughout this work, we applied the entire procedure to two additional gas samples: $N_2O$ (non-symmetric) and $CS_2$ (symmetric). The experimental and theoretical results are given in the section 3 of the supplementary information.

### Conclusions

In this paper we explored and demonstrated the ramifications of the l-Doubling to the alignment dynamics of non-symmetric (OCS) and symmetric ($CO_2$) triatomic molecular models. The results highlight the importance of l-Doubling for correct quantification of rotational decay phenomena, especially at the initial times after rotational excitation even at room temperature and more severely at higher temperatures. The l-Doubling is typically overshadowed by collisional decay already at sub-atmospheric gas pressures, making it an elusive phenomenon that can be easily overlooked. However its great importance to current research of nonsecular collisions and rotational decay in general emanates from its potential to 'imitate' decaying dynamics that results merely from interference of rotational signals associated with the ground and thermally excited bending modes. As a purely coherent phenomenon, l-Doubling can be deciphered from collision-induced decay via echo spectroscopy.

**Acknowledgments:** We thank Mr. Eran Rosen and the TAU chemistry machine shop team for producing the spectroscopic gas cell and the graphics of Fig.1.

**Funding:** The authors acknowledge the support of the Israel Science Foundation (1856/22) and the PAZI foundation.

# Supplementary Information

## 1) Table for constants

Values are given in units of $[cm^{-1}]$

| Molecule | $\omega_{bend}$ | $B_{000}$ | $B_{010}^+$ | $B_{010}^-$ | $q$ | $D$ |
|---|---|---|---|---|---|---|
| $CO_2$ | 667 | 0.3902 | 0.3912 | 0.3905 | $65 \cdot 10^{-5}$ | $12 \cdot 10^{-8}$ |
| $OCS^{32}$ | 539 | 0.20285 | 0.20331 | 0.2031 | $21 \cdot 10^{-5}$ | $4.3422 \cdot 10^{-8}$ |
| $N_2O$ | 588.7 | 0.41901 | 0.41997 | 0.41917 | $79 \cdot 10^{-5}$ | $17.6 \cdot 10^{-8}$ |
| $CS_2$ | 398.4 | 0.1091 | 0.10935 | 0.10927 | $7.5 \cdot 10^{-5}$ | $0.993 \cdot 10^{-8}$ |

Table 1: summarizing table of vibrational bending frequencies and rotational constants for $CO_2$[44], $OCS$[36], $N_2O$[45] and $CS_2$[46,47] molecules. Reference papers are given in brackets.

## 2) The effect of excitation pulse width

The l-doubling dynamics demonstrated in OCS and $CO_2$ naturally depends on the different rotational constants associated with the ground and excited bending modes. These rotational constants ($B_{000}, B_{010}^+, B_{010}^-$) dictate the revival periods, i.e. the times at which the three selective alignment signals will temporally overlap and interfere. The exact outcome of this three-signal interference, strongly depends on their specific shape and duration. While the shape of the alignment transient is primarily dictated by the quantum rotational dynamics, the duration of the transients can controlled by varying the duration of the excitation pulse. The interplay between the two time-scales, i.e. the (fixed) revival periodicity and the (controlled) duration of the interfering transients provides additional means to control the l-doubling temporal dynamics. In what follows, we exemplify the outcomes of this interplay by the simulations presented in figure S1. We simulate the rotational dynamics of $CO_2$ induced by varying excitation pulse durations. As the pulse duration increases (from 100fs to 2ps), the duration of the revival transients elongates respectively and modulation observed in the l-doubling dynamics stretches in time respectively.

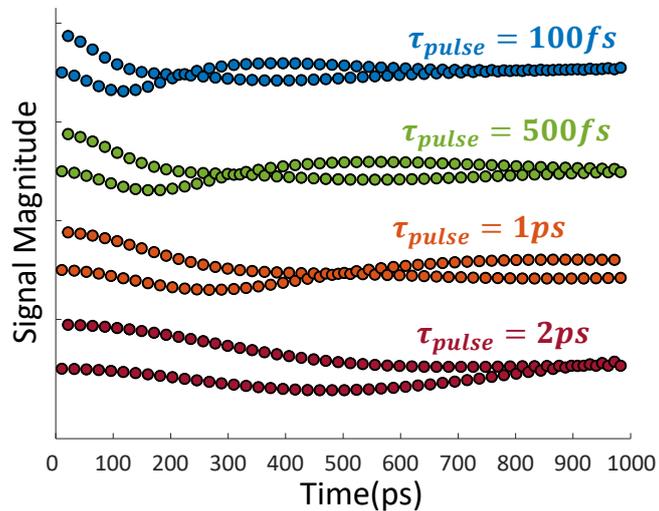

Figure S1: Simulated revival magnitudes for $CO_2$ for different excitation pulse durations: 0.1ps, 0.5ps, 1ps, and 2ps.

This is readily observed by considering the time at which the halves and quarter revival magnitudes cross each other.

## 3) l-Doubling dynamics in other gas samples: $N_2O$ and $CS_2$

To further validate our analysis strategy, we present the experimental overlaid with theoretical results of two additional gas species: $N_2O$ (non-symmetric) and $CS_2$ (symmetric). The analysis strategy, starting from the raw optical-birefringence measurements followed by quantification of the signal magnitudes and the simulation technique is identical to that elaborated in the main paper file. Note that modulation observed in $N_2O$ takes a different shape compared to that of OCS, and emanates from the specific values of the rotational coefficients (see table 1) and their interferences throughout the periodic revivals of $N_2O$.

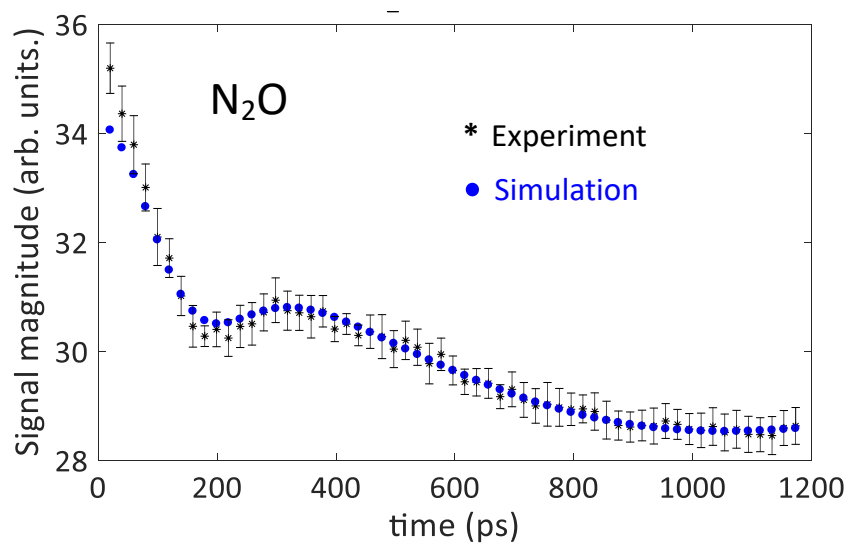

Figure S2: l-Doubling dynamics measured in 10torr $N_2O$ sample at 333K.

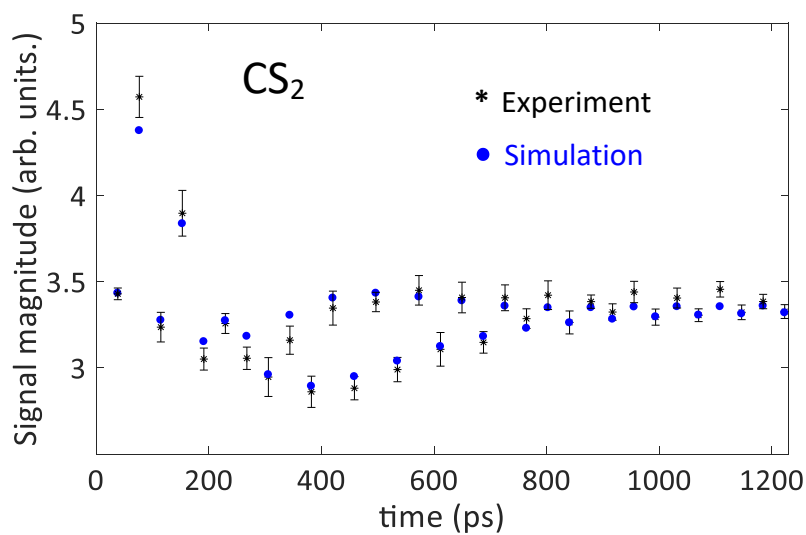

Figure S3: l-Doubling dynamics measured in 7torr $CS_2$ sample at 323K.